\documentclass[a4paper]{article}

\usepackage{INTERSPEECH2020}
\usepackage{url}
\usepackage{hyperref}

\usepackage[utf8]{inputenc}

\title{Target-Speaker Voice Activity Detection: a Novel Approach for Multi-Speaker Diarization in a Dinner Party Scenario}
\name{Ivan Medennikov$^{1,2}$, Maxim Korenevsky$^1$, Tatiana Prisyach$^1$, Yuri Khokhlov$^1$, \\ Mariya Korenevskaya$^1$, Ivan Sorokin$^1$, Tatiana Timofeeva$^1$, Anton Mitrofanov$^1$, \\ Andrei Andrusenko$^{2}$, Ivan Podluzhny$^1$, Aleksandr Laptev$^{2}$, Aleksei Romanenko$^{1,2}$}
\address{$^1$STC-innovations Ltd, St. Petersburg, Russia \quad $^2$ITMO University, St. Petersburg, Russia}
\email{ \{medennikov, korenevsky, knyazeva, khokhlov, korenevskaya, sorokin, timofeeva,\\ mitrofanov-aa, andrusenko, podluzhnyi, laptev, romanenko\}@speechpro.com}

\begin{document}

\maketitle

\begin{abstract}
Speaker diarization for real-life scenarios is an extremely challenging problem. Widely used clustering-based diarization approaches perform rather poorly in such conditions, mainly due to the limited ability to handle overlapping speech. We propose a novel Target-Speaker Voice Activity Detection (TS-VAD) approach, which directly predicts an activity of each speaker on each time frame. TS-VAD model takes conventional speech features (e.g., MFCC) along with i-vectors for each speaker as inputs. A set of binary classification output layers produces activities of each speaker. I-vectors can be estimated iteratively, starting with a strong clustering-based diarization.

We also extend the TS-VAD approach to the multi-microphone case using a simple attention mechanism on top of hidden representations extracted from the single-channel TS-VAD model.
Moreover, post-processing strategies for the predicted speaker activity probabilities are investigated.
Experiments on the CHiME-6 unsegmented data show that TS-VAD achieves state-of-the-art results outperforming the baseline x-vector-based system by more than 30\% Diarization Error Rate (DER) abs.

\end{abstract}
\noindent\textbf{Index Terms}: speaker diarization, TS-VAD, CHiME-6

\section{Introduction}
\label{sec:intro}
Diarization is a process of determining boundaries of utterances for each speaker in a conversation. Diarization is an important part of many applications, primarily of automatic speech recognition (ASR), e.g., for meeting minutes creation. A conventional approach \cite{Sell_2018, Diez_2018} consists of several stages, namely speech/voice activity detection (SAD/VAD), segmentation of the detected speech into short subsegments, and extraction of the current speaker's features (i-vectors \cite{Dehak_2011}, d-vectors  \cite{Variani_2014, Wan_2017}, x-vectors\cite{Snyder_2018} etc.) followed by clustering (k-means \cite{Dimitriadis_2017}, agglomerative hierarchical \cite{Garcia-Romero_2017}, spectral \cite{Ning_2006}, etc.) according to some similarity metrics (Probabilistic Linear Discriminant Analysis~(PLDA) \cite{Prince_2007, Garcia-Romero_2017} score, cosine, etc.). These stages can also be followed by re-segmentation (such as GMM \cite{Fredouille_2009}, Variational Bayes  \cite{Sell_2015} or LSTM-based \cite{Sahidullah_2019}) and overlapping speech segments post-processing.

Currently, high diarization accuracy is achieved for many benchmarks, such as CallHome.
However, the development of a diarization system for complex acoustic environments is still an unsolved task.
This was a motivation for the DIHARD Challenges \cite{Ryant_2018, Ryant_2019}
focused on the development of systems for ``hard'' diarization. The DIHARD II Challenge  \cite{Ryant_2019} includes, in particular, multichannel Tracks 3 and 4 based on the CHiME-5 Challenge \cite{Barker_2018} data, which are very hard for both diarization and ASR. The same data diarization is also one of the CHiME-6 Challenge~\cite{Watanabe_2020}
Track 2 tasks. This data recorded  in real-life conditions contains a large amount of overlapping speech. Conventional diarization systems are not well-suited for processing highly overlapping speech, so it is not very surprising that even the best DIHARD II system developed by BUT \cite{Landini_2020} provided DER only slightly below 60\%.

During the participation in the CHiME-6 Challenge, we were solving the same diarization problem~\cite{stc_chime6}. Achieving high diarization accuracy was crucial for high ASR performance, which was the main challenge goal. So we started by reviewing approaches that are effective in diarizing highly overlapping speech. One of the most promising methods of such kind is the end-to-end neural diarization (EEND) \cite{Fujita_2019}, which performs diarization in a single stage and outputs the frame-level activity probabilities for each speaker independently. Another direction that we found to be promising consists of using pre-computed features of a speaker of interest to draw the system's attention to only their speech. This direction is represented by such approaches as Target-Speaker ASR \cite{Kanda_2019[b]}, Speaker Beam \cite{Zmolikova_2017, Delcroix_2018} and Voice Filter \cite{Wang_2018} aimed at the target-speaker speech extraction, etc. Moreover, the TS-ASR approach may be used for simultaneous speech recognition and diarization \cite{Kanda_2019[c]}. One more representative of this direction is the Personal VAD \cite{Ding_2019} approach allowing to detect speech boundaries for only that speaker whose acoustic ``profile'' is fed into the system.

Inspired by the the ideas mentioned above, we combined their benefits in our own diarization approach for the CHiME-6 referred to as Target-Speaker VAD (TS-VAD).
In the course of the development, our model evolved from the simple one, very similar to the Personal VAD, to the sophisticated one, which processes multi-channel recordings and outputs independent speaker activity streams like the EEND model. The main difficulty for the successful TS-VAD application is that, unlike TS-ASR or Personal VAD scenarios, we do not have any pre-computed speakers' features and have to compute them directly from severely distorted and highly overlapping speech. Nevertheless, we managed to find such a way of applying TS-VAD which reduced DER to the values of 33\% and 36\% for the CHiME-6 development and evaluation sets respectively, which is much better than BUT system result for DIHARD II as well as the CHiME-6 baseline result. Thus, although the direct comparison to the DIHARD II results may not be fully correct\footnote{See Section \ref{sec:chime} for the detailed explanation.}, the obtained results show the potential of our approach for diarization in CHiME-6-like scenarios. Our TS-VAD implementation is available as a part of a new Kaldi recipe for CHiME-6\footnote{https://github.com/kaldi-asr/kaldi/tree/master/egs/chime6/s5b\_track2}. 

The rest of the paper is organized as follows: in Section \ref{sec:chime} we describe in more details the CHiME-6 Challenge conditions and data, then in Section \ref{sec:single} we talk about how a single-channel version of TS-VAD evolved and how the above-mentioned prototypes influenced this process. The multi-channel TS-VAD system is described in Section \ref{sec:multi}, and the fusion of different systems as well as post-processing of diarization results are discussed in Section \ref{sec:fusion}. Section \ref{sec:conclusions} provides some conclusions and suggests possible future directions.

\section{CHiME-6 Challenge}
\label{sec:chime}
The CHiME-6 Challenge continues a series of challenges on speech recognition in complex real-life acoustic environments. It is based on the same data as the previous CHiME-5, which consists of multi-channel recordings from six 4-microphone Microsoft Kinect arrays, located in three different rooms. Each recording contains an informal conversation of four persons in a dinner party scenario in three locations. Besides, the two-channel recordings from in-ear microphone pairs worn by each person are also provided for the training. Data includes 20 sessions (16 for training, 2 for development and 2 for evaluation).
The goal is to create ASR system for multi-channel Kinect recordings in two tracks. In Track 1 participants are allowed to use manual boundaries of each speaker's utterances created manually from worn microphones audio listening, while in Track 2 this information can not be used. Since the knowledge of boundaries can help to improve ASR substantially, one of the Track 2 tasks is to obtain a high-quality diarization of Kinect recordings. This task is very similar to Track 4 of the DIHARD II Challenge.
Same as in DIHARD II, the metrics for this task are DER and Jaccard Error Rate~(JER), which are evaluated without using a collar and without excluding overlapping segments. Nonetheless, there are some differences, as well.
Firstly, CHiME-6 organizers provided software for time-aligning different channels within a session, which was not available in DIHARD II. All our results are obtained on pre-aligned data. Also, unlike DIHARD II allowing participants to use any training data, 
the CHiME-6 rules limit training data for Track 2 by only CHiME-6 data and VoxCeleb data. 

Initially, reference rttm-files provided by the CHiME-6 organizers were created based on utterance boundaries set manually (the same information was used in Tracks 3-4 of DIHARD II). 
However, such segmentation contains intra-utterance pauses, which are treated as speech, and unlabelled speakers' introductions. Therefore, the organizers provided another reference segmentation with the exclusion of silence segments. This was carried out using triphone GMM-HMM forced alignment of reference transcripts over manual segments.
Besides, the UEM-file was provided, which makes it possible to exclude speaker introductions from scoring. The significant difference between reference rttm files in CHiME-6 and DIHARD II (about 30-40\% in terms of DER) is the main reason why the direct comparison of the results is not possible. 

CHiME-6 data is rather noisy and reverberated as well as contains a significant amount of overlapping speech, so it is difficult to perform an accurate diarization with clustering-based systems. 
Statistics of overlapping speech in the development and evaluation datasets is shown in Table \ref{tab:overlap}. 
If one changes the reference rttms to leave only a single speaker on each overlapping segment, this will result in DER of 25.61\%/21.76\%  (due to the miss errors only) for the development and evaluation sets, respectively. These values show the lowest achievable DERs on the CHiME-6 data for clustering-based diarization systems without special overlaps processing.

\begin{table}[!ht]
\vspace{-2mm}
    \centering
    \begin{tabular}{lccccc}
      & 0 & 1 & 2 & 3 & 4 \\
      \midrule
      dev     & 24.05\% & 54.25\% & 17.74\% & 3.49\% & 0.47\% \\
      eval    & 33.47\% & 51.52\% & 12.03\% & 2.47\% & 0.51\%
    \end{tabular}
    \caption{Distribution of audio with respect to the number of simultaneously speaking persons.}
    \label{tab:overlap}
\vspace{-8mm}
\end{table}

\section{Single-channel TS-VAD}
\label{sec:single}

\subsection{Single-Speaker model}

Recently, a series of papers on multi-speaker speech processing was published, where models focus on a specific speaker ignoring the speech of others. These approaches include TS-ASR \cite{Kanda_2019[b]} for target speech recognition, Speaker Beam \cite{Zmolikova_2017, Delcroix_2018} and Voice Filter \cite{Wang_2018} for target speech extraction, and Personal VAD \cite{Ding_2019} for target speech detection. Most of them use the acoustic footprint of a target speaker (usually i-vector), obtained during prior enrollment, to focus on speech of interest. The Personal VAD model seemed to be most appropriate for our goals since it selects each speaker's speech independently. However, this model was trained and tested on concatenated instead of overlapped speech segments, so it was unclear if it could handle overlapping speech as well. Besides, under the CHiME-6 conditions, it was not trivial how to compute i-vectors for each speaker since the use of manual segmentation was prohibited in test time.

Thus, we started with a purely research question: given an i-vector estimated on manual non-overlapping speech segments of a target speaker, is it possible to detect their speech in overlapping conditions? The answer was generally positive. Our first single-speaker TS-VAD model was very similar to the Personal VAD with the same three targets obtained from the forced alignment, namely silence, target speech, and non-target speech. The model architecture was a simple 3-layer BLSTM with projections~\cite{Sak_2014}.

All the experiments were performed in the Kaldi ASR Toolkit~\cite{Povey_2011}. We used the acoustic model training dataset and i-vector extractor from the Kaldi \textit{chime6} recipe.  Moreover, we used a ``negative'' version of each utterance with an i-vector corresponding to a random speaker from the same session and device (without such ``negative'' examples, the model tends to detect any speech as a target). Training targets were taken from the \textit{tri3} GMM forced alignment: \textit{silence} and \textit{noise} phones were treated as silence class, and the rest phones as target speech or non-target speech for the original and ``negative'' utterances, respectively. 

TS-VAD model outputs a sequence of probabilities of target speaker presence for each time frame. To convert it into the segmentation, the simple post-processing described in Section~\ref{sec:fusion} was applied.
After multiple experiments with i-vectors computed on non-overlapping regions of manual segmentation, we managed to obtain DER=66.81\% on the development data (for all experiments on manual segmentation, we used CH1 of the reference device, i.e., the device closest to the currently active speaker).

Single-speaker TS-VAD processes each speaker independently, which is intuitively sub-optimal. So, we introduced an additional ``mutual'' threshold to the post-processing procedure. Considering a set of speech probabilities on the current frame, we found that there is likely no speech of those speakers whose probabilities are strongly dominated by the maximum probability on this frame. Thus, the probability was set to zero if it differed from the maximum probability on the current frame by more than the threshold.
This trick provided a dramatic diarization improvement reducing DER to 46.12\%, which looked promising.

\subsection{Multi-Speaker model}
We also experimented with the EEND model \cite{Fujita_2019}, hoping it would work well for CHiME-6 data. Unfortunately, this was not the case due to several reasons.
First of all, the relatively short audio pieces are required to apply EEND since the self-attention module needs to see the whole sequence. But such block-wise processing induces the between-block speaker permutation problem.
We used oracle permutation to estimate the lowest possible DER for the EEND.
An oracle DER of the EEND model trained on the VoxCeleb data was comparable to DER of the baseline diarization, presumably due to the severe acoustic mismatch between the VoxCeleb and CHiME-6 data. 
The model trained on CHiME-6 data provided much better oracle DER (about 50\%).
To resolve between-block permutations, we extracted baseline x-vectors from the EEND-induced segments and applied several clustering approaches to them. We found that the simple unconstrained k-means algorithm worked best, but the obtained DERs were only slightly better 
than those of the baseline diarization.
Nonetheless, we adopted some ideas from EEND to improve the TS-VAD model.

\begin{figure}[ht]
\vspace{-4mm}
  \centering
  \includegraphics[width=200pt]{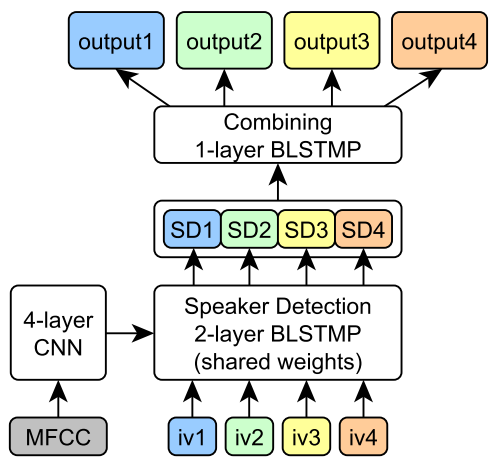}
\vspace{-4mm}
  \caption{Single-channel TS-VAD scheme}
  \label{fig:TS-VAD}
\vspace{-4mm}
\end{figure}

As the CHiME-6 diarization task is closed (the number of speakers in a session is always four), we decided to use i-vectors of all speakers in the session as inputs, instead of the target speaker's i-vector only. The initial model was designed to estimate speech probability for the speaker corresponding to the first i-vector. It was beneficial to average the predicted probabilities obtained on various permutations of i-vectors.
However, much better results were obtained with the TS-VAD model designed to predict speech probabilities for all speakers simultaneously, as the EEND model does. This model with four output layers was trained using a sum of binary cross-entropies as a loss function. We also found that it is essential to process each speaker by the same Speaker Detection~(SD) 2-layer BLSTMP, and then combine SD outputs for all speakers by one more BLSTMP layer. The model architecture is shown in Figure~\ref{fig:TS-VAD}.
Note that parameters of the SD block are shared across speakers, and it is trained jointly with the whole TS-VAD model.

As we performed all the experiments in the Kaldi ASR Toolkit~\cite{Povey_2011}, it was easier to use 2-class softmax instead of a sigmoid in the output layers.
Training targets were 8-dimensional vectors representing four pairs of silence and speech probabilities corresponding to four speakers.
Given an utterance, targets corresponding to the current speaker were taken directly from the forced alignment. Targets for the three other speakers were obtained by averaging alignments from neighboring overlapping utterances over all the devices and channels (for non-overlapping frames, targets for speech and silence were 0 and 1, respectively).

To provide more data variability, we performed on-the-fly random permutations of speakers (i.e., both i-vectors and targets) during the training. It was also beneficial to use the mixup data augmentation~\cite{orig_mixup} using our Kaldi-compatible tools presented in~\cite{stc_mixup}.
Moreover, we obtained a small DER improvement (about 0.5\%) by adding an 800h subset of the VoxCeleb data augmented by artificial room impulse responses.
The final multi-speaker model achieved an impressive DER of 37.40\% using i-vectors estimated on manual segmentation.

Note that the ``mutual'' threshold providing a huge DER reduction for the single-speaker TS-VAD turned out to be totally useless for the multi-speaker model.

\subsection{Estimation of i-vectors}

After the experiments on manual segments, we switched to the real CHiME-6 task and tried to compute i-vectors from the segmentation provided by the baseline clustering-based diarization (DER=63.42\% on the development set). 

Unfortunately, the early versions of TS-VAD did not provide any tangible improvement over the baseline results.
So, we firstly focused on improving the baseline diarization. To this end, we applied the improved x-vectors extractor based on a Wide ResNet~(WRN), which was trained on the VoxCeleb data. The details on the extractor are given in~\cite{Gusev_2020}. Application of the baseline Agglomerative Hierarchical Clustering (AHC) based on 
PLDA scoring to these x-vectors improved DER by about 10\% abs. Then we used the idea from \cite{Park_2020} and replaced AHC clustering based on PLDA scores with Spectral Clustering~(SC) based on Cosine Similarities. Using similarity matrix binarization with respect to automatically chosen threshold \cite{Park_2020} provided the consistent DER improvement on both the development and evaluation sets (see Table \ref{tab:track2_results_diar} for the results). Besides, using WRN x-vectors substantially improved the permutation resolution on the EEND segmentation and covered a large fraction of the gap to the oracle DER.

The improved clustering-based diarization provided a good initial estimation of i-vectors for TS-VAD. The next idea was to re-estimate i-vectors iteratively, using segmentation from the previous TS-VAD iteration. However, we found that even better results can be achieved using probabilities obtained by the TS-VAD as soft-weights for re-estimation of i-vectors. Note that, to ensure robust i-vectors estimation, for each speaker we considered only those frames where speech probability was more than 0.8 of total speech probability for all speakers on a given frame. The second iteration provided a significant gain, but the third one did not lead to any improvement. Note that such i-vectors estimation provides as good results as i-vectors computed on manual segmentation\footnote{In the same conditions without WPE and channel averaging, iterative i-vector estimation led to 38\% DER, compared to 37.4\% on the manual segments.}.
Later we found that the same iterative procedure with the best TS-VAD model starting from the baseline diarization also improves DER by about 15\% absolute. However, more iterations are required for convergence. 

\section{Multi-channel processing}
\label{sec:multi}
The single-channel version of TS-VAD~(TS-VAD-1C) processes each channel separately, but we suggested that the multi-channel processing may be beneficial. Indeed, we found that the multi-channel Weighted Prediction Error~(WPE) dereverberation~\cite{Yoshioka_2012,Drude_2018} improves the results of TS-VAD by about 1\% DER absolute. Moreover, averaging of per-channel TS-VAD probabilities provides up to 2\% absolute DER reduction. 

\begin{figure}[ht]
\vspace{-4mm}
  \centering
  \includegraphics[width=200pt]{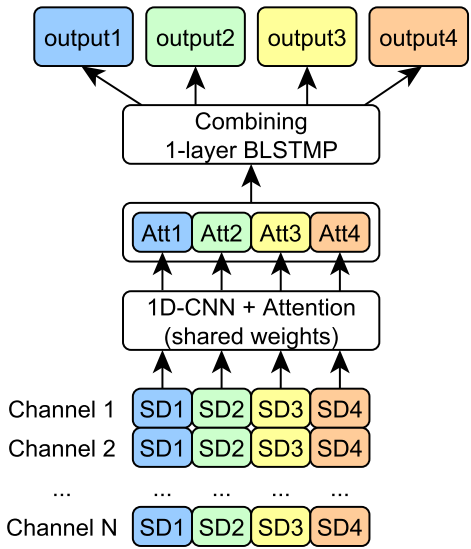}
\vspace{-4mm}
  \caption{Multi-channel TS-VAD scheme}
  \label{fig:TS-VAD-MC}
\vspace{-3mm}
\end{figure}

To process separate Kinect channels jointly, we investigated the multi-channel TS-VAD model (TS-VAD-MC), which takes a combination of TS-VAD-1C model SD blocks outputs from a set of 10 Kinect recordings as an input. The architecture of this model is presented in Figure~\ref{fig:TS-VAD-MC}. The channels of input Kinect recordings are chosen randomly for training, while CH1 and CH4 are taken at test-time. This way of combining information from different channels is more effective than a simple averaging of probabilities, as in the TS-VAD-1C model. All the SD vectors for each speaker are passed through a 1-d convolutional layer and then combined by means of a simple attention mechanism. Combined outputs of attention for all speakers are passed through a single BLSTM layer and converted into a set of per-frame probabilities of each speaker's presence/absence.

Finally, to improve overall diarization performance, we fused 3 single-channel and 3 multi-channel TS-VAD models by computing a weighted average of their probability streams. Fusion weights were tuned to minimize DER on the development set. The diarization results on different stages of our system are presented in Table \ref{tab:track2_results_diar}. 

\begin{table}[!ht]
\vspace{-3mm}
    \centering
    \begin{tabular}{l|c|c|c|c}
                          & \multicolumn{2}{c|}{DEV} & \multicolumn{2}{c}{EVAL} \\
                             & DER   &  JER    & DER   & JER     \\
    \midrule
    x-vectors + AHC          & 63.42 & 70.83   & 68.20 &  72.54  \\
    EEND + WRN x-vectors     & 52.20 & 57.42   & 56.01 &  61.49  \\
    WRN x-vectors + AHC      & 53.45 & 56.76   & 63.79 &  62.02  \\
    WRN x-vectors + SC       & 47.29 & 49.03   & 60.10 &  57.99  \\
    \quad + TS-VAD-1C (it1)  & 39.19 & 40.87   & 45.01 &  47.03  \\
    \qquad + TS-VAD-1C (it2) & 35.80 & 37.38   & 39.80 &  41.79  \\
    \qquad + TS-VAD-MC       & 34.59 & 36.73   & 37.57 &  40.51  \\
    \midrule
    Fusion                   & \textbf{32.84} & \textbf{36.31}   & \textbf{36.02} &  \textbf{40.10}  \\
    Fusion*                  & 41.76 & 44.04   & 40.71 &  45.32  \\
    \end{tabular}
    \caption{Diarization results (* stands for DIHARD II reference) }
    \label{tab:track2_results_diar}
\vspace{-11mm}
\end{table}

\section{Post-processing}
\label{sec:fusion}
To convert the TS-VAD output probabilities into a sequence of segments, the simple post-processing was applied.
It includes 51-tap median filtering, binarization with the threshold of 0.4, combining speech segments separated by pauses shorter than 0.3s, and deleting speech segments shorter than 0.2s.

Alternatively, Viterbi decoding was applied for the post-processing. We introduced a simple Hidden Markov Model (HMM) with 11 states representing silence, speech from each of four speakers without overlaps, and overlapping speech from 6 possible pairs of speakers (overlaps of three and four speakers were neglected due to short duration). The emission probabilities in each state were induced from TS-VAD output probabilities, while the transition probabilities were tuned to minimize DER on the development set. The transitions from the silence state to the two-speaker states and vice versa were prohibited and other transition probabilities were shared between any pairs of states with the same number of speakers (7 tunable parameters in total). The most likely state sequence found by the Viterbi search determined the final segments. 

The influence of several post-processing techniques on DER improvement is shown in Table \ref{tab:fusion}.
\begin{table}[!ht]
 \vspace{-2mm}
   \centering
    \begin{tabular}{l|c|c|c}
               & Post-processing & {DEV} & {EVAL} \\
    \midrule
    Best single TS-VAD &  T+F+S     & 34.59   & 37.57   \\
    Fusion       & T            & 34.73   & 37.52  \\
    Fusion       & T+F+S        & 33.56   & 36.63  \\
    Fusion       & V+S          & 32.84   & 36.02  \\
     \end{tabular}
    \caption{DER for different post-processing. T, F, S and V stand for binarization with the threshold, median filtering, short speech/pause processing, and Viterbi decoding, respectively.}
    \label{tab:fusion}
\vspace{-10mm}
\end{table}

\section{Conclusions}
\label{sec:conclusions}
We presented a novel approach for the diarization of multi-speaker conversations, which provided state-of-the-art results in a complex multi-channel dinner party scenario. The proposed Target-Speaker VAD selects speech of every conversation participant taking their i-vector as an input along with MFCC features. Since the sufficiently good segmentation is required for the reliable i-vectors estimation, we improved the baseline clustering-based diarization significantly. 

It is worth noting that we also tried to replace i-vectors with more discriminative speaker embeddings like x-vectors as the TS-VAD inputs, but results got much worse. We believe the reason is that the relatively simple TS-VAD architecture is unable to match well the current MFCC features to the speaker embeddings obtained from a much more complicated model. Another possible reason is severe overfitting due to a sparse embeddings space and a small number of speakers in the training data.

Although our final solution is task-dependent (multi-channel input, a fixed number of speakers), we believe the proposed approach is flexible enough to be easily modified for other similar tasks. In the future, we plan to extend our solution to the scenario of informal meetings with an unknown large number of participants.

\section{Acknowledgements}
This work was partially financially supported by the Government of the Russian Federation (Grant 08-08).

We are grateful to STC Voice Biometrics Team for the awesome speaker embeddings extractor and valuable discussions.

\bibliographystyle{IEEEtran_2020}

\bibliography{mybib}
\end{document}